\documentclass[aps,pre,twocolumn,groupedaddress,floatfix]{revtex4}
\usepackage{epsfig}
\newcommand{\mon}{\begin{displaymath}}
\newcommand{\moff}{\end{displaymath}}
\renewcommand{\bf}[1]{\mbox{\boldmath ${#1}$}}
\renewcommand{\b}[1]{\vec{#1}}
\newcommand{\pd}[2]{\frac{\partial {#1}}{\partial {#2}}}
\newcommand{\od}[2]{\frac{d {#1}}{d {#2}}}
\newcommand{\eon}{\begin{equation}}
\newcommand{\eoff}{\end{equation}}
\newcommand{\eq}[1]{Eq. (\ref{#1})}

\newcommand{\del}{\nabla}
\newcommand{\mat}[1]{\bf{\mathcal{#1}}}
\newcommand{\fig}[1]{figure \ref{#1}}
\newcommand{\twovect}[2]{\left( \begin{array}{c} {#1} \\ {#2} \end{array} \right) }
\newcommand{\chem}[2]{{}^{#2} {#1}}

\begin{document}
\title{A Quasispecies on a Moving Oasis}
\author{Michael M. Desai}
\email{desai@fas.harvard.edu}
\affiliation{Department of Physics, Harvard University,
Cambridge MA 02138}
\author{David R. Nelson}
\affiliation{Department of
Physics, Harvard University, Cambridge MA 02138}
\date{\today}

\begin{abstract}
A population evolving in an inhomogeneous environment will adapt
differently to different regions.  We study the conditions under
which such a population can maintain adaptations to a particular
region when that region is not stationary, but can move.  In
particular, we study a quasispecies living near a favorable patch
(``oasis'') in the middle of a large ``desert.'' The population
has two genetic states, one of which which conveys a relative
advantage while in the oasis at the cost of a disadvantage in the
desert. We consider the population dynamics when the oasis is
moving, or equivalently some form of ``wind'' is blowing the
population away from the oasis.  We find that the ratio of the two
types of individuals exhibits sharp transitions at particular
oasis velocities.  We calculate an extinction velocity, and a
switching velocity above which the dominance switches from the
oasis-adapted genotype to the desert-adapted one. This switching
velocity is analagous to the quasispecies mutational error
threshold.  Above this velocity, the population cannot maintain
adaptations to the properties of the oasis.
\end{abstract}

\maketitle

Spatial inhomogeneities in the environment are often essential to
understanding the dynamics of natural populations.  Populations
may for example be confined to limited reserves or live in an
environment with gradients in resources or large-scale
inhomogeneities in habitability.  The evolution of a population in
an inhomogeneous environment is particularly interesting.  When
individuals move over the environment sufficiently rapidly
relative to the length scale of the inhomogeneities, they may all
see and adapt to an ``averaged'' environment.  When this is not
true, however, the dynamics can be much more complex.  Some
individuals may randomly see primarily one part of the range
while others see other parts.  The population in the distant
future will likely be dominated by the descendents of a few
exceptionally lucky individuals in the present, who will have seen
an unusually favorable subset of the set of possible environments.
Thus it is not at all clear a priori exactly how different regions
of the environment will influence the evolution.

In this paper, we consider a simple environment with two regions,
a favorable ``oasis'' in a large less favorable (or unfavorable)
``desert.'' The favorable area could be realized as a game
reserve, a region of favorable climactic conditions, or a patch of
light, among other things.  Provided that the favorable region is
stationary and sufficiently large, a population will adapt to the
special properties of this region.  However, in many cases the
oasis will move at some typical speed $v$, for example due to
seasonal weather patterns or human intervention.  (Equivalently,
the population could be blown across the oasis at speed $v$ by
some form of ``wind'').  In this situation, there is a velocity
beyond which the population will not be able to maintain
adaptations to the properties of the oasis.  Rather, the evolution
will be dominated only by the desert environment.  We focus in
this paper on calculating this velocity, which we call the
``switching velocity.''

Recent work has considered the dynamics, though not the evolution,
of a population being blown by some form of convective ``wind''
across a nonuniform environment.  In these models, the equation
for the population density at a point $(\b{x}, t)$ is \eon
\pd{c(\b{x}, t)}{t} = D \del^2 c - \b{v} \cdot \del c + r(\b{x})
c, \eoff where $D$ is the diffusion constant, $\b{v}$ is the drift
velocity, and $r(\b{x})$ is the growth rate.  This equation thus
describes a population multiplying in response to some spatially
heterogeneous environment while diffusing and drifting.  It can be
analyzed using techniques adapted from an analysis of
non-Hermitian Schroedinger-like operators \cite{252, 255}.  A
nonlinear saturation term can be added, and is important for some
purposes.

Ref. \cite{111} examines this model in two dimensions in the case where
$r(\b{x})$ is random, with only short-range correlations.  In the limit of
small $v$ ($v \ll v_F \equiv 2 \sqrt{a D}$, where $a$ is the average of
$r(x)$), the population is dominated by a few colonies that grow up around
``hot spots,'' regions which happen to have higher growth rates than
surrounding areas.  As $v$ increases, individual colonies on less prosperous
hot spots are blown away in a series of ``delocalization'' transitions.  In
the limit of large $v$ all organisms are blown away from individual hot spots
and convect across the environment.  At long times, the population is
dominated by those individuals that happened to take very special paths
through the random environment, travelling through a disproportionate number
of the hot spots.  Thus this is an example of a system in which a typical
individual will evolve in response to a very special subset of the
environment.

In subsequent work, Dahmen \emph{et. al.} examined the transition between the
small and large $v$ regimes by looking at a model of a single ``hot spot''
\cite{230}.  These authors took the growth rate $r(\b{x})$ to be large within
a small hot spot, or ``oasis,'' and smaller (and possibly negative) in the
surrounding ``desert.'' They analyzed the dynamics near the extinction and
delocalization velocities, where the population is blown off of the oasis.
Their predictions have been qualitatively confirmed by recent experiments on
\emph{Bacillus subtilis} growth \cite{231}.

In this paper, we extend this work to consider the dynamics of a population
of multiple types of individuals in a desert-oasis environment.  In
particular, we consider a quasispecies model with two types of individuals.
Thus $c(\b{x},t)$ becomes a two-vector \eon \bf{c}(\b{x}, t) = \left(
\begin{array}{c} c_{1}(\b{x}, t) \\ c_{2}(\b{x}, t)  \end{array} \right),
\eoff and $r(\b{x})$ becomes a two-by-two matrix with diagonal
elements specifying the growth rates of the two genotypes and
off-diagonal elements specifying the mutation and back-mutation
rates.  Genotype $1$ represents the population of some ``ideal''
genome sequence, while type $2$ represents all other less ideal
sequences.  There is mutation back and forth between the two,
typically with a mutation rate away from the ideal sequence greater
than the mutation rate towards it.  Without any spatial variation, (i.e. with
$r(x)$ independent of $x$), this is a simple quasispecies model. A
few recent papers have examined other versions of a spatial
quasispecies model, with many different possible types of
individuals representing all possible Hamming distances from the
ideal sequence \cite{245, 227}.  Here, we focus on a simple
two-type spatial model, although it is straightforward to
generalize our results.

We assume that the population $c_{1}$ has some gene or sequence which conveys
an advantage in the environment of the oasis, but has an overall cost which
makes it disadvantageous in the desert.  The population $c_{2}$ consists of
all individuals who have lost the function of the gene by one or more
mutations.  Individuals of type $1$ are thus those that have adapted to the
special properties of the oasis, while type $2$ individuals are
desert-adapted.

We examine the population dynamics as we change the velocity $v$.
We find two important transitions.  When the growth rate in the
desert is negative, there is an extinction velocity $v_e$ where
the entire population goes extinct.  When the genotype $2$ desert
growth rate is positive, there is a ``switching'' velocity $v_s$
where the behavior of the population changes dramatically. Below
this velocity the population (particularly within the oasis) is
dominated by genotype $1$.  Above it, the oasis-adapted genotype
is outcompeted, genotype $2$ dominates, and $c_1/c_2 \to 0$ at
long times. This transition is a velocity-driven analog of the
classical mutational error threshold in non-spatial quasispecies
models \cite{altmeyerref1, altmeyerref2, altmeyerref5}.

These results have interesting implications.  If the oasis
represents some part of a species' habitat, the switching velocity
is simply how quickly this can move before the species can no
longer maintain adaptations to the properties of this aspect of
its range.  Our analysis is also a first step towards
understanding the spatial quasispecies model in a random
environment, where the population does not evolve in response to
the averaged environment but rather in response to a different,
highly selective subset of the environment.

The outline of this paper is as follows.  In section \ref{model}
we give a detailed description of our model.  In section
\ref{analysis} we outline the calculation of the critical
velocities and the behavior of the populations in the different
regimes.  In section \ref{simulation} we compare our analytical
results with computer simulations.  Finally, in section
\ref{discussion} we discuss the main biological implications of
these calculations.

\section{Model} \label{model}
We consider a population with two types of individuals whose
densities are described by \eon \bf{c}(\b{x}, t) = \left(
\begin{array}{c} c_{1}(\b{x}, t) \\ c_{2}(\b{x}, t) \end{array}
\right). \eoff The dynamics is exponential growth (or decay) with
diffusion and convection \eon \pd{\bf{c}(\b{x}, t)}{t} = D
\del^{2} \bf{c} - \b{v} \cdot \b{\del} \bf{c} + \mat{R} (\b{x})
\bf{c}, \label{eq4} \eoff where $\mat{R} (\b{x})$ is a two-by-two
matrix. We are primarily concerned with the extinction and
delocalization transitions.  Thus for the bulk of our analysis, we
neglect the possibility of a nonlinear saturation term.  In
section \ref{nonlinear} we discuss the consequences of such a
nonlinearity.

We define the growth and death rates of individuals of type $i$
within the oasis to be $\alpha_{i}$ and $\gamma_{i}$ respectively.
Outside the oasis, the growth and death rates are defined to be
$\beta_{i}$ and $\delta_{i}$. The mutation rate from type 1 to
type 2 is defined to be $\mu_{1}$, and the back-mutation rate is
$\mu_{2}$.  Based on these definitions, the matrix $\mat{R}
(\b{x})$ is given by \eon \mat{R}(\b{x}) = \left(
\begin{array}{cc} \alpha_{1} - \gamma_{1} - \mu_{1} \alpha_{1} &
\mu_{2} \alpha_{2}
\\ \mu_{1} \alpha_{1} & \alpha_{2} - \gamma_{2} - \mu_{2}
\alpha_{2}
\end{array} \right) \eoff inside the oasis and similarly (with
$\alpha \to \beta$ and $\gamma \to \delta) $ in the desert.

For simplicity, we assume that the growth rates within the desert
and the oasis are the same, i.e. $\alpha_{i} = \beta_{i}$.  The
desert is less hospitable because of a higher death rate, not a
lower growth rate.  This simplifies the analysis but is not a
serious limitation;  the results for $\alpha \neq \beta$ are
straightfoward to calculate by the same methods. We next define
$a_{i} \equiv \alpha_{1} - \gamma_1$ and $b_{i} \equiv \beta_{i} -
\delta_{i}$.  For convenience we also assume that $a_1 = \alpha_1$
and $a_2 = \alpha_2$ (i.e. $\gamma_1 = \gamma_2 = 0$).  It is easy
to relax this assumption, but the results more transparent with it
in place. These simplifications yield the matrix \eon \mat{R}
(\b{x}) = \left( \begin{array}{cc} a_{1} - \mu_{1} a_{1} & \mu_{2} a_{2} \\
\mu_{1} a_{1} & a_{2} - \mu_{2} a_{2} \end{array} \right) \qquad
\textrm{(inside oasis)} \eoff in the oasis and similarly in the
desert, \eon \mat{R} (\b{x}) = \left(
\begin{array}{cc} b_{1} - \mu_{1} a_{1} & \mu_{2} a_{2} \\ \mu_{1}
a_{1} & b_{2} - \mu_{2} a_{2} \end{array} \right) \qquad
\textrm{(outside oasis)}. \eoff

We assume that $a_{1} > b_{1}$ and $a_{2} > b_{2}$, so that we
have a beneficial oasis and a harmful desert.  We further assume
that $a_{1} > a_{2}$ and $b_{2} > b_{1}$ (which also implies
$a_{2} > b_{1}$) so that genotype $1$ is better in the oasis and
genotype $2$ is better in the desert. Absent these inequalities, one or
the other type will unequivocally dominate the population (up to
the mutational error threshold) independent of the drift velocity.
In that case, we can ignore the inferior type at long times and
the problem reduces to that studied in \cite{230}.  Our
assumptions imply that individuals of type 1 have some function
that conveys an advantage inside the oasis, at the cost of a
disadvantage elsewhere.  Our analysis will determine whether or
not this function can be maintained in the face of mutational
pressure when the oasis is moving at velocity $v$.

\begin{figure}
\begin{center}
\epsfig{file=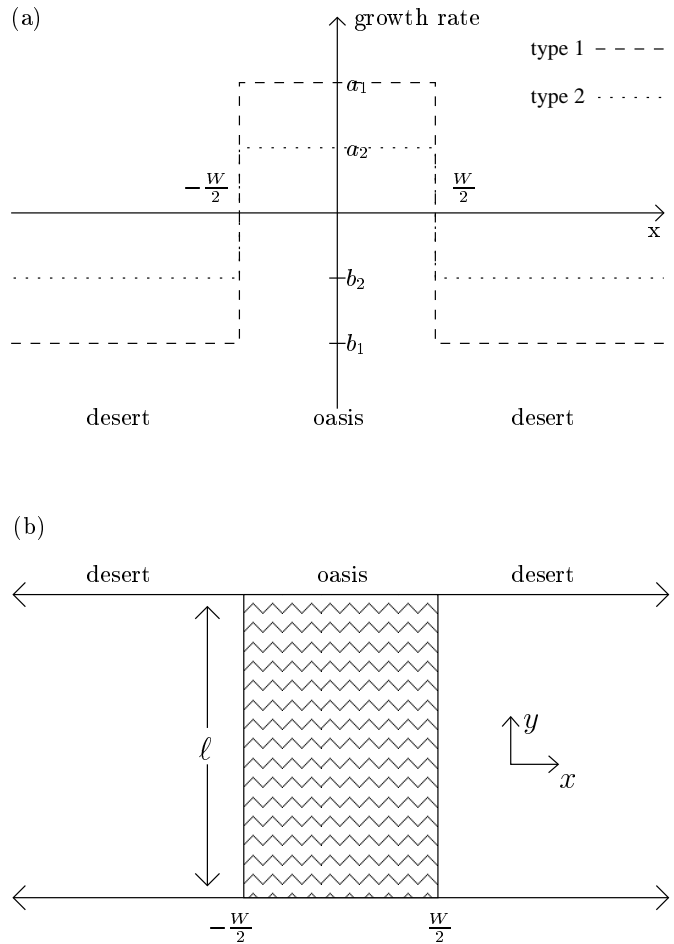, width=3.5in} \caption{(a) shows the
one-dimensional system we consider, for the case of a deadly
desert.  The dashed line is the growth rate of genotype 1, the
dotted line the growth rate of genotype 2.  This is an
approximation to the two-dimensional system shown in (b), valid
when the initial conditions are uniform in the $y$ direction or
the time is greater than $\ell^2 / D$.  \label{fig1}}
\end{center}
\end{figure}

We could of course make any number of assumptions about the
geometry of the desert and the oasis.  We will consider for
simplicity a one-dimensional system with an oasis of width $W$ in
the middle of an infinite desert, as depicted in figure 1a.  This
analysis also describes the long time behavior of the
two-dimensional system with geometry as described by figure 1b. If
the initial conditions are uniform in the $y$ direction, the
one-dimensional system of figure 1a is equivalent to the
two-dimensional system of figure 1b. If the conditions are
nonuniform, the two become equivalent after a time of order
$\ell^2/D$ \cite{230}. More complicated geometries are certainly
possible, and we discuss these briefly in section
\ref{geometries}.

We assume throughout that differential equations are an adequate
representation of these biological processes.  We neglect
discreteness in population number, which may have some importance
near the extinction transition.  The effects of discreteness may
be analyzed using the methods of \cite{46, 54, 56, 58, 59, 60,
61}.

\section{Calculation of the Population Dynamics} \label{analysis}
We analyze our model using the non-Hermitian methods of Refs.
\cite{252, 255, 111}.  We first rewrite our system in the form
\eon \pd{\bf{c} (x, t)}{t} = \mat{L} \bf{c}, \eoff where $\mat{L}$
is given by $\mat{L} = \mat{L}_{in} \theta ( W - |x| ) +
\mat{L}_{out} \theta (|x| - W)$, and $W$ is the width of the
oasis. The function $\theta (y) = 1$ if $y \geq 0$ and $0$
otherwise, and {\small \begin{eqnarray} \mat{L}_{in} & = & \left(
\begin{array}{cc} D
\partial_x^2 - v \partial_x +a_1 - \mu_1 a_1 & \mu_{2} a_{2} \\
\mu_{1} a_{1} & D \partial_x^2 - v \partial_x + a_2 - \mu_2 a_2
\end{array} \right), \nonumber \\
\mat{L}_{out} & = & \left( \begin{array}{cc} D \partial_x^2 - v \partial_x
+ b_1 - \mu_1 a_1 &
\mu_{2}
a_{2} \\ \mu_{1} a_{1} & D \partial_x^2 - v \partial_x + b_2 - \mu_2 a_2
\end{array} \right) .  \nonumber \end{eqnarray} }
$\mat{L}$ is non-Hermitian, but we can diagonalize it with a
system of left and right eigenfunctions $\bf{\phi}_n^R(x)$ and
$\bf{\phi}_n^L(x)$, with their (common) eigenvalues $\Gamma_n$.
These eigenfunctions (which we also call ``states'') satisfy the
orthonormality condition \eon \int \bf{\phi}_m^{L} (x) \cdot
\bf{\phi}_n^R (x) dx = \delta_{mn}. \eoff Using this result, we
can write the initial condition as a linear superposition of the
right eigenfunctions, i.e. $\bf{c} (x, t=0) = \sum_n c_n
\bf{\phi}_n^R (x)$, where \eon c_n = \int d^d x \bf{\phi}_n^{L}
(x) \cdot \bf{c}(x, t=0). \eoff We can then immediately write down
the solution valid for all times, namely \eon \bf{c}(x, t) =
\sum_n c_n \bf{\phi}_n^R (x) e^{\Gamma_n t}. \label{eqnsolution}
\eoff

This result has a clear biological interpretation.  Initially, the
population is in some particular arrangement, expressible as a
linear combination of the different eigenfunctions. This
combination will necessarily include the eigenfunction
corresponding to the largest $\Gamma_{n} \equiv \Gamma_{gs}$
(which is always real)  \cite{endnote2}. This eigenfunction is
special, and we refer to it as the ``ground state''
\cite{endnote4}.  As time passes, the population distribution
looks more and more like the ground state.  This distribution will
grow (or die out) exponentially with rate $\Gamma_{gs}$.  The
other $\bf{\phi}_{m}^R$ may be important initially, but since they
grow more slowly than the ground state, they soon become
irrelevant.  Thus understanding the ground state function and its
eigenvalue are the key to understanding the long time behavior of
the population.

It remains to solve for the eigenfunctions and eigenvalues
$\bf{\phi}_n^L(x), \bf{\phi}_n^R(x)$, and $\Gamma_n$.  For the
case $v = 0$, the solution is straightforward and will be
discussed in detail below.  For nonzero $v$, we make use of the
fact that the eigenfunctions of $\mat{L}_{v = 0}$ are related to
the eigenfunctions of $\mat{L}$ by an ``imaginary gauge
transformation'' \cite{111}.  That is, if $\bf{\phi}_{n,
v=0}^R(x)$ is a right eigenfunction of $\mat{L}_{v=0}$ with
eigenvalue $\Gamma_n$, then \eon \bf{\phi}^R_{n, v}(x) = e^{vx/2D}
\bf{\phi}_{n, v=0}^R(x) \eoff is a right eigenfunction of
$\mat{L}$, with eigenvalue \eon \Gamma_n^v = \Gamma_n^{v=0} -
\frac{v^2}{4D}, \eoff as can be verified by allowing $\mat{L}$ to
act on $\bf{\phi}_n^R$. Similar expressions hold for
$\bf{\phi}^L_{n,v}$.  The eigenvalues $\Gamma_n^{v = 0}$ are all
real.  Thus, eigenfunctions for $v > 0$ are very similar to the
eigenfunctions for $v = 0$.  The genotype 1 versus genotype 2
composition of the states is not altered at all.  The only change
is that the wind causes a distortion of the population in the
direction of the wind, and the growth rates of the states shift
downward ``rigidly'' (i.e. independent of $n$) by an amount
$\frac{v^2}{4D}$.

However, this procedure works only for small $v$.  To see this,
consider the behavior of the eigenfunctions as $x \to \infty$.  We
expect (and will soon verify) that the $v = 0$ eigenfunctions far
from the oasis decay exponentially, \eon \phi_{n, v=0} \sim
e^{-\kappa_n |x|}. \eoff Thus for $v > 0$, \eon \phi_n^R \sim
e^{vx/2D - \kappa_n |x|}. \eoff When $v < 2D \kappa_n$, the
eigenfunctions vanish at infinity, as they should.  However, for
$v > 2D \kappa_n$, this function blows up at infinity, which is
unreasonable.  The correct eigenfunctions have a different
character when $v > 2D \kappa_n \equiv v^*_n$.

\begin{figure}
\begin{center}
\epsfig{file=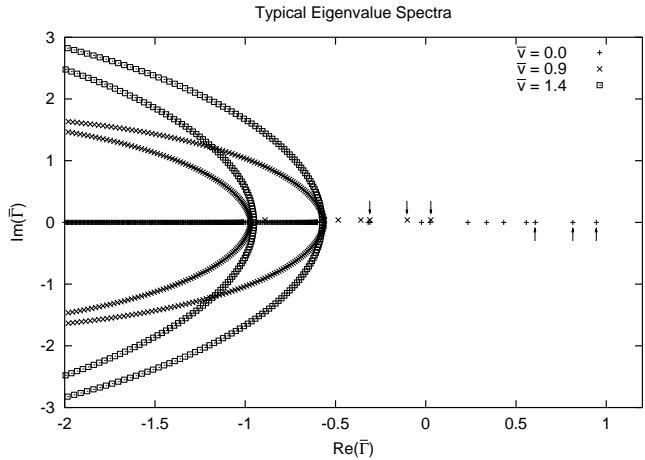,width=3.5in} \caption{An example of
eigenvalue spectra, shown here for $3$ values of $\bar v =
\frac{v}{v_F} \equiv \frac{v}{2 \sqrt{D a_1}}$ with $\bar a_1 =1,
\bar b_1 = -1, \bar a_2 = 0.6, \bar b_2 = -0.6, \mu_1 = 0.01,
\mu_2 = 0.001$, and $\bar D = 0.25$, where the overbars indicate
the non-dimensionalized parameters discussed in Appendix C. To
allow easy distinction from the $\bar v = 0$ results, the
localized $\bar v = 0.9$ eigenvalues have been shifted slightly
upwards. Note the rigid shift of the localized eigenvalues (those
on the real axis) to the left as we increase $\bar v$.  This is
highlighted by the three $\bar v = 0$ states marked by upwards
arrows, which for $\bar v = 0.9$ are shifted into the three states
marked by downards arrows. The two (and only two) delocalization
transitions $\bar \Gamma_1^* \approx -1$ and $\bar \Gamma_2^*
\approx -0.6$ are also clearly visible. Note that these transition
points, and all the delocalized states, do \emph{not} shift to the
left as $\bar v$ increases.  On the left half of the spectrum
(omitted from the figure), the delocalized eigenstates form closed
loops, an artifact of the computational discretization used to
produce this figure. \label{fig2}}
\end{center}
\end{figure}

Ref. \cite{252} shows that the transition at $v_n^*$ is a
delocalization of the corresponding eigenfunction.  For $v <
v^*_n$, the eigenfunctions are localized around the oasis, but for
$v > v^*_n$ they are delocalized.  This makes intuitive sense. For
small velocities, the population will tend to cluster around the
oasis, but for larger wind velocities it gets blown off the oasis
and must live by drifting across the desert.  The behavior of the
eigenvalues $\Gamma_{n}$ near this delocalization transition is
striking. Up to $v^*_n$, the eigenvalue $\Gamma_n^v$ is simply
equal to $\Gamma_n^{v=0} - \frac{v^2}{4D}$, but beyond this point
this relation no longer holds. Instead, $\Gamma_n$ jumps off the
real axis at $v^*_n$, becoming complex, and the eigenfunctions
become broad delocalized states extending through the desert
\cite{252, 255, 111}.  We denote the value of $\Gamma_n$ at which
this occurs by $\Gamma^*_n$.  As we continue to increase $v$ above
$v_n^*$, the real part of $\Gamma_n$ stays approximately constant,
although the imaginary part does change.  From the gauge
transformation relationship, we have $\Gamma_n^* = \Gamma^{v=0}_n
- \frac{(v^*_n)^2}{4D}$.  However, as we will see, the structure
of the $n$-dependence of $v_n$ and $\Gamma_n$ is such that there
are only two different values of $\Gamma_n^*$. This is a crucial
point.  In our problem, the states will divide into those
dominated by genotype 1 and those dominated by genotype 2.  As we
will show, states dominated by genotype $1$ delocalize at
$\Gamma_1^* = \langle r_1 \rangle$, the spatial
average growth rate of the first genotype, which up to finite size
effects is just $b_1$. $\Gamma_2^* = \langle r_2 \rangle \approx
b_2$ plays the same role for states dominated by the second
genotype. By our assumptions about the parameters, $\Gamma_2^* >
\Gamma_1^*$.  An example of eigenvalue spectra for several
values of $v$ is given in \fig{fig2}.

\begin{figure}
\begin{center}
\epsfig{file=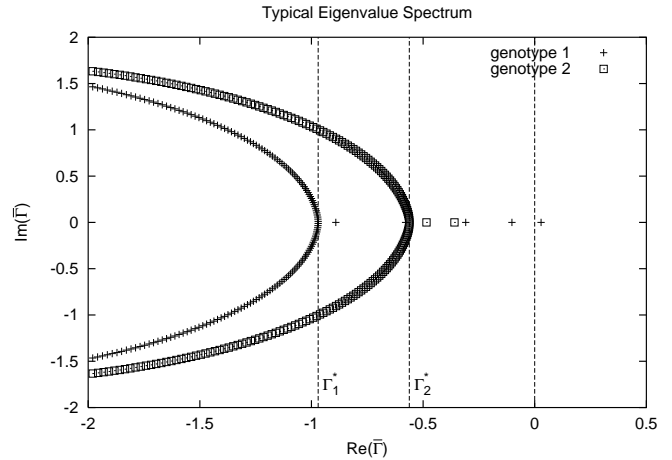,width=3.5in} \caption{A single eigenvalue
spectrum, with the genotype-1 and genotype-2 dominated states
distinguished. This is just the $\bar v = 0.9$ spectrum from
\fig{fig2} above. Note the state with the largest Re$(\Gamma)$
(i.e. the ground state) is a genotype-1 dominated state.  This
$\bar v$ is just below the extinction velocity, so $\Gamma_{gs}$
is just above the extinction threshhold, indicated by the dashed
vertical line at $\Gamma = 0$. As $\bar v$ increases, all of the
localized (real) states move to the left until they enter either
the left parabola of delocalized states (for genotype-1 dominated
states) or the right parabola of delocalized states (for
genotype-2 dominated states).  Thus as $\bar v$ increases
$\Gamma_{gs}$ will soon pass through $0$ at the ``extinction
velocity''.  However, the delocalized states do not shift to the
left as $\bar v$ increases. Thus as $\bar v$ increases further,
$\Gamma_{gs}$ will eventually pass through $\Gamma_2^*$, indicated
by the center vertical line. Once this happens, the largest
Re$(\Gamma)$ no longer belongs to the original genotype-1
dominated state, but rather to a genotype-2 dominated delocalized
state.  The critical velocity at which this occurs is the
``switching velocity.''  In this example, since $\Gamma_2^* < 0$,
the switching velocity is higher than the extinction velocity, and
is therefore biological uninteresting. However, if we had
$\Gamma_2^*
> 0$, the spectrum would look identical, except it would be
shifted to the right.  We would then have $\Gamma_2^* > 0$, and
hence there would always be states above Re$(\Gamma) = 0$. Thus
there would be no extinction velocity, and the switching velocity
would be biologically relevant. \label{fig22}}
\end{center}
\end{figure}

Each localized state has some particular $\Gamma_n^{v=0}$, and for
nonzero $v$ its eigenvalue becomes $\Gamma_n = \Gamma_n^{v = 0} -
\frac{v^2}{4D}$. As $v$ increases, $\Gamma$ decreases until the
state delocalizes, at $\Gamma_1^*$ for those states dominated by
type 1 and at $\Gamma_2^*$ for those states dominated by type 2.
We will see that the ground state for $v = 0$ is dominated by
genotype 1.  As $v$ increases there comes a critical point when
this ground state eigenvalue $\Gamma_{gs}$ crosses $\Gamma_2^*$.
Beyond this point it is no longer the ground state.  Rather, the
delocalized genotype-2 dominated state with eigenvalue
$\Gamma_2^*$ has the highest $\Gamma$, and this state determines
the population dynamics.  We call this critical velocity the
``switching velocity,'' where the dominance switches from genotype
1 to genotype 2.  This switch is a type of quasispecies
transition, caused not by exceeding a mutation rate error
threshold, but by exceeding a critical velocity.  This reasoning
is illustrated in \fig{fig22}.

If the growth rate for genotype 2 in the desert is negative, then
$\Gamma_2^* < 0$.  The switching behavior will then be difficult
to observe experimentally, as both genotypes will be going extinct
when it happens.  In this case, the biologically interesting
transition occurs at the ``extinction'' velocity where the growth
rate of the ground state passes through $0$.  Beyond this
velocity, all the states have negative growth rate, so neither
genotype can survive. When the growth rate for genotype 2 in the
desert ($b_2$) is positive, the switching velocity becomes
biologically relevant. In this case, there is no extinction
velocity.  This is because the delocalized states
do not shift to lower $\Gamma$ as $v$ increases, and at least
one genotype-2 dominated delocalized state has eigenvalue
$\Gamma_2^* \approx b_2 > 0$. Intuitively, this makes sense
because the population, dominated by genotype 2, can survive in
the desert at arbitrarily large velocities.

\subsection{Solution for the Eigenvalues and Eigenfunctions}
In order to carry out the analysis sketched above, we must solve
for the $v = 0$ eigenvalues and eigenfunctions, and determine the
values of $\Gamma_1^*$ and $\Gamma_2^*$.  It is possible to do
this exactly, and an outline of this calculation is presented in
Appendix A.  However, it is more straightforward and instructive
to first examine the solutions for $\mu_1 = \mu_2 = 0$, and then
use perturbation theory to find the results for small $\mu_1$ and
$\mu_2$.

\subsubsection{The $\mu_1 = \mu_2 = 0$ Solution} \label{mu0}
We first examine the system for $\mu_1 = \mu_2 = 0$.  In this
case, the two types of individuals are completely independent. The
problem reduces to that studied in \cite{230}.  We use the
imaginary gauge transformation to eliminate the velocity, and
focus on the $v = 0$ eigenvalues and eigenstates.  We then have to
solve the eigenvalue equation \eon \mat{L}_{v=0} \phi_n^{i, v=0} =
\Gamma_n^{i, v=0} \phi_n^{i, v=0}. \eoff  This problem is formally
equivalent to the square well problem in quantum mechanics
\cite{qmbook}. The $v = 0$ right eigenstates are of the form \eon
\bf{\phi}_n^{1, v=0} = \left(
\begin{array}{c} \psi_n^{1} \\ 0 \end{array} \right), \qquad
\bf{\phi}_n^{2, v=0} = \left(
\begin{array}{c} 0 \\ \psi_n^{2} \end{array} \right), \eoff where
\mon \psi_n^i = \left\{ \begin{array}{ll} A_n^i e^{\kappa_n^i x} &
\mathrm{for} \quad x < - W/2 \\ B_n^i e^{i k_n^i x} + C_n^i e^{-i
k_n^i x} & \mathrm{for } \quad - W/2 < x < W/2 \\ G_n^i e^{-
\kappa_n^i x} & \mathrm{for } \quad x > W/2.
\end{array} \right. \moff
Note that the index $i$ indicates the genotype that dominates the
state.  Substituting this ansatz into the eigenvalue equation
leads to \eon \Gamma_n^{i, v=0} = D (\kappa_n^i)^2 + b_i = - D
(k_n^i)^2 + a_i. \label{gammak} \eoff

We proceed by requiring that $\bf{\phi}$ and its first derivative
be continuous at $x = \pm W/2$, which determines the constants $A,
B, C$, and $G$ up to an overall normalization and yields a
transcendental equation to determine $\Gamma$.  This analysis is
carried out in detail in \cite{230}.  The essential result is that
provided the oasis is wide enough that a typical individual gives
birth many times while diffusing across it, the ground state
eigenvalue can be approximated as $\Gamma^{v=0, \mu =0}_{gs}
\approx a_1$, with a corresponding eigenfunction that is entirely
genotype 1 and localized largely within the oasis.  We will use
this approximation throughout the rest of this paper
\cite{endnote3}. We can also calculate the position of the
delocalization transitions $\Gamma_n^{i*}$.  These are defined as
the amount by which the $v=0$ eigenvalue $\Gamma_n^{i, v=0}$ is
shifted by the delocalization velocity $v_n^{i*}$. We thus have
$\Gamma_n^{i*} = \Gamma_n^{i, v=0} - \frac{(v_n^{i*})^2}{4D} =
\Gamma_n^{i, v=0} - D (\kappa_n^{i})^2$. Using \eq{gammak}, we
find $\Gamma_n^{i*} = b_i$.  Note that, as claimed above, this
$\Gamma_n^{i*}$ is independent of $n$ and depends only on $i$.
Thus there are two and only two delocalization threshholds, one
corresponding to each genotype.

\subsubsection{Perturbation Theory in $\mu_1$ and $\mu_2$}
We can now examine the results for nonzero mutation rates $\mu_1,
\mu_2 > 0$ by using a non-Hermitian version of time-independent
perturbation theory from quantum mechanics \cite{qmbook}.  We
require that $\mu_1$ and $\mu_2$ be small, specifically that
$\frac{\mu_i a_i}{a_1 - a_2}, \frac{\mu_i a_i}{b_2 - b_1} \ll 1$.
The details of the calculation are described in Appendix B.  The
eigenstates now all involve both genotypes, although those that
began as genotype $1$ states remain dominated by this type, and
vice versa.  More precisely, we have
\eon \bf{\phi}_n^{1, v=0, R} = \left( \begin{array}{c} \psi_n^1 \\
\frac{\mu_1 a_1}{a_1 - a_2} \psi_n^2 \end{array} \right), \eoff
for a genotype-1 dominated localized state, and \eon
\bf{\phi}_n^{2, v=0, R} = \left( \begin{array}{c} \frac{\mu_2
a_2}{b_2 - b_1} \psi_n^1 \\ \psi_n^2 \end{array} \right) \eoff for
a genotype-2 dominated delocalized state, where $\psi_n^1$ and
$\psi_n^2$ are given in section \ref{mu0}.  Note that we focus on
genotype-1 dominated localized states and genotype-2 dominated
delocalized states because no other state can dominate the
dynamics in any regime.  The eigenvalues are also shifted. The
eigenvalues for genotype-1 dominated localized states become \eon
\Gamma_n^{1, v=0} = \Gamma_n^{i, v = 0, \mu = 0} - \mu_1 a_1 +
\frac{\mu_1 a_1 \mu_2 a_2}{a_1 - a_2}, \label{eq24} \eoff while
the genotype-2 dominated delocalized states have eigenvalues \eon
\Gamma_n^{2, v=0} = \Gamma_n^{2, v=0, \mu=0} - \mu_2 a_2 +
\frac{\mu_1 a_1 \mu_2 a_2}{b_2 - b_1}, \eoff plus higher order
terms in $\mu_1$ and $\mu_2$.

\subsection{Critical Velocities}
We can now calculate the critical velocities at which the dynamics
changes qualitatively.  There are two relevant cases.  For $b_2 <
0$ neither delocalization transition occurs at positive $\Gamma$
because neither genotype can survive in the desert.  Thus the
switching velocity, though it exists formally, is biologically
irrelevant.  The extinction velocity $v_e$ is the velocity where
the ground state eigenvalue passes through zero.  Below $v_e$ a
genotype-1 dominated population can multiply but above this
velocity the population must go extinct.  This velocity is defined
by $\Gamma_{gs}^{v_e} = \Gamma_{gs}^{v = 0} - \frac{v_e^2}{4D} =
0$, which using \eq{eq24} gives \eon v_e = 2 \sqrt{(D a_1)} \left[
1 - \frac{\mu_1}{2} \right], \eoff valid to first order in $\mu_1$
and $\mu_2$.  Note that for $\mu_1 = 0$, we have $v_e = v_F \equiv
2 \sqrt{D a_1}$, the Fisher velocity for genotype $1$.

For $b_2 > 0$ there is no extinction velocity, as genotype 2 can
survive at any velocity.  However, there is a switching velocity
$v_s$ where the population shifts from being mostly genotype 1 to
mostly genotype 2. This occurs when the growth rate of the
genotype-1 dominated ground state passes through $\Gamma_2^* = b_2
- \mu_2 a_2 + \frac{(\mu_2 a_2)^2}{b_2 - b_1}$.  By a similar
calculation we find \eon \label{switching} v_s = 2 \sqrt{D (a_1 -
b_2)} \left[ 1 - \frac{1}{2} \frac{\mu_1 a_1}{a_1 - b_2} +
\frac{1}{2} \frac{\mu_2 a_2}{a_1 - b_2} \right], \eoff also valid
to first order in $\mu_1$ and $\mu_2$.  As mentioned above, this
switching velocity is a sort of quasispecies transition.  As the
velocity increases passes this critical threshold, the population
can no longer maintain the ``ideal'' sequence, even if it is below
the mutational error threshold.  This switching velocity is also
well-defined for $b_2 < 0$, but will be difficult to observe
because $v_s > v_e$.  We can easily calculate the second-order
corrections to both $v_e$ and $v_s$ (see Appendix B).

\begin{figure}
\begin{center}
\epsfig{file=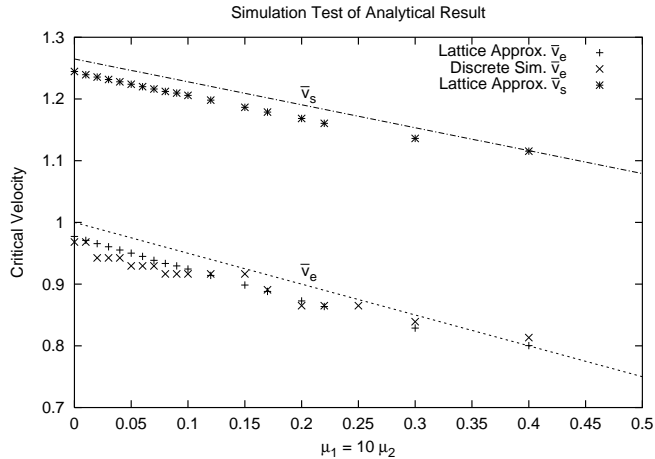,width=3.5in} \caption{A comparison of the
critical velocities $\bar v_e$ and $\bar v_s$ from the simulations
to the analytical result, as a function of $\mu_1 = 10 \mu_2$. The
analytical result is shown as a dashed line for $\bar v_s$ and a
dotted line for $\bar v_e$.  Here we have $\bar a_1 = 1$, $\bar
a_2 = 0.6$, $\bar b_1 = -1$, $\bar b_2 = -0.6$, and $\bar D =
0.25$, where the overbars indicate the dimensionless units
described in Appendix C. This result is typical of such
comparisons for other parameter values. For these parameters, we
expect the perturbation calculation of the analytical result to be
valid for $\mu_1 \ll 0.4$.  The slight overestimates of $\bar v_e$
and $\bar v_s$ by the analytic theory for small $\mu_1$ are due to
finite size effects. The underestimate as $\mu_1$ grows beyond the
range of our perturbation expansion is due to the importance of
higher-order terms. \label{fig3}}
\end{center}
\end{figure}

\section{Simulations} \label{simulation}
We use two different computational methods to test our analytic
results. First, we use a lattice discretization of the Liouville
operator $\mat{L}$ to calculate the eigenvalue and eigenfunction
spectrum for particular sets of parameter values.  This numerical
work provides an aid to intuition and one check of the validity of
our approximations. Second, we simulate the underlying discrete
process, namely individuals multiplying and mutating at
appropriate rates.  This tests not just the results of our
analysis, but also the overall applicability of continuous time
differential equations to the real discrete populations we model.

\subsection{Lattice Approximation to the Liouville Operator}
We made several approximations in arriving at our analytical
results, including an approximations for $\Gamma_{gs}$ and for the
shifts in eigenvalues with $\mu_1$ and $\mu_2$.  To test these
approximations, we calculate the eigenvalue spectrum numerically.
We discretize space, find the resulting discretized Liouville
operator, and numerically diagonalize it for a particular set of
parameters.  The details of this method are described in Appendix
C.

This approach allows us to determine the shifts in the growth
rates as $\mu$ or $v$ is varied, and hence the critical velocities
$v_e$ and $v_s$.  We can compare these results to our analytical
predictions, and therefore confirm our calculation of the critical
velocities.  This comparison for one particular set of parameters
is shown in \fig{fig3}. Comparisons for other parameter values
have been carried out, and give similar results.  The eigenvalue
spectra shown in \fig{fig2} and \fig{fig22} were also obtained in
this way.

\subsection{Simulations of the Discrete System}
We can also simulate the underlying discrete process which
inspired our formulation of the differential equations we analyze
in this paper.  We discretize space, placing individuals on a
one-dimensional lattice which represents our environment.  These
individuals move around, proliferate, die, and mutate.  We also
impose a saturation term so that at long times the population
distribution settles down to a steady state.

By comparing the steady state population profiles for different values of
$v$, we determine $v_e$ or $v_s$.  These results can then be compared to the
analytical results.  This comparison, for one particular set of parameters,
is shown in \fig{fig3}.  Note that this comparison is only for $v_e$, as for
these parameter values $v_s$ is not biologically relevant and is thus
impossible to observe with this type of simulation.  The details of this
method are described in Appendix D.

\section{Discussion} \label{discussion}
To interpret our results, it is helpful to first consider the behavior of
a non-spatial quasispecies model.  We imagine a population living in a
uniform environment whose conditions match those of the oasis.  Genotype 1
grows more quickly, but there is a mutational pressure away from this
``ideal'' genotype.  Mutation away from this sequence is typically
expected to be much more frequent than mutation back, so it is common in
such models to set $\mu_2 = 0$.  It is then straightforward to calculate
the composition of the population.  We find that the equilibrium ratio
of genotype $1$ to genotype $2$ individuals $q \equiv c_1/c_2$ is given by
\eon q = \frac{(1 - \mu_1) a_1 - a_2}{\mu_1 a_1}. \eoff Thus the ratio of
species $1$ to $2$ decreases with increasing $\mu_1$ until $\mu_1$ reaches
the critical ``error threshold'' for this quasispecies model.  At this
threshold, $\mu_1^c = \frac{a_1 - a_2}{a_1}$, the ``ideal'' genotype can
no longer survive and the population becomes completely dominated by
genotype $2$.  This is the most famous result of Eigen's quasispecies model
\cite{altmeyerref1, altmeyerref2, altmeyerref5}.

\begin{figure}
\begin{center}
\epsfig{file=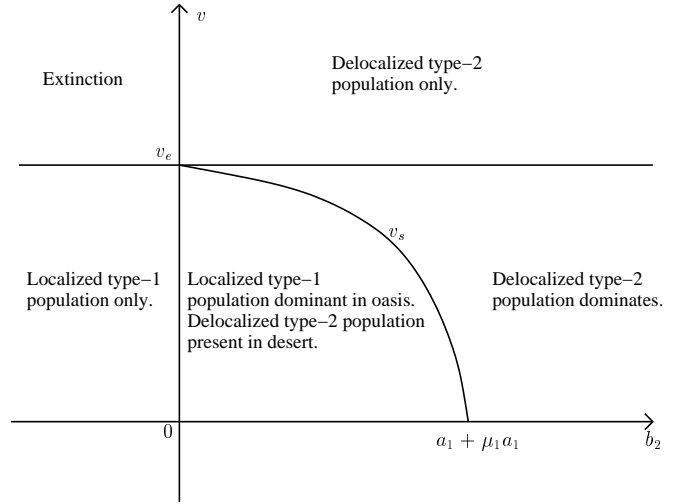, width=3.5in} \caption{A phase diagram
showing the transitions as a function of $v$ and $b_2$. This
figure accounts for the qualitative effects of a nonlinear
saturation term.  \label{fig4}}
\end{center}
\end{figure}

Our analysis finds that, in analogy to the quasispecies error
threshold, in the spatial model there is a \emph{velocity}
threshold above which genotype $1$ is outcompeted by genotype $2$.
This velocity, which we call the switching velocity $v_s$, is
given by \eq{switching}.  If the growth rate of genotype 2 in the
desert is positive, we can expect to see this behavior in a real
system. Our model naturally also has the traditional error
threshold; for $\mu_1 > \frac{a_1 - a_2}{a_1}$ and $\mu_2 = 0$,
genotype $1$ is outcompeted by genotype $2$ regardless of $v$. It
would be interesting to explore the interactions between the
mutation-driven and velocity-driven quasispecies transitions.
However, our analysis is based on the assumption of small mutation
rates and thus focuses on the velocity-driven transition under the
assumption that we are well away from the mutation-driven
transition \cite{endnote1}.  A qualitative phase diagram of the
different velocity-driven transitions is given in \fig{fig4}.

Our model describes a number of important biological situations.
The oasis could be a particularly favorable patch of the
environment or a zone of favorable climactic conditions which is
moving at a typical speed $v$ due to seasonal weather patterns or
shifts in climate.  A population can take advantage of this oasis
by adapting to these conditions, but then will fare worse in the
rest of the space.  Our analysis explores how fast the oasis can
move before the population can no longer maintain an adaptation to
the favorable patch.  Beyond this speed individuals may
occasionally find themselves in the oasis but cannot adapt quickly
enough to benefit.

Besides showing the existence of the velocity-driven quasispecies
transition, our results make $5$ interesting biological
predictions.  We discuss each of these in turn.

\emph{(1) The overall population growth rate decreases as $v^2$ up
to the critical velocity.} The imaginary gauge transformation
tells us that for $b_2 < 0$, the exponential growth rate of the
population decreases as $\frac{v^2}{4D}$ until it goes extinct at
$v_e = 2 \sqrt{a_1 D} \left( 1 - \frac{\mu_1}{2} \right)$.  For
$b_2 > 0$ the exponential growth rate of the population decreases
again by $\frac{v^2}{4D}$ until it reaches $v_s = 2 \sqrt{(a_1 -
b_2) D} \left( 1 - \frac{\mu_1 a_1}{a_1 - b_2} + \frac{\mu_2
a_2}{a_1 - b_2} \right)$.  Increasing the velocity further does
not change the overall growth rate because a delocalized genotype
$2$ population then dominates.

\emph{(2) Below the critical velocity, the genotype-1 dominated
population will extend somewhat into the desert.} The type-1
population will extend a typical distance $\xi =
\frac{1}{\kappa_{gs} - \frac{v}{2D}}$ into the desert.  From the
relationship between $\kappa$ and $\Gamma$ we find that this
typical distance is \eon \xi = \frac{2D}{2 \sqrt{D (a_1 - b_1)} -
v}, \eoff independent of the mutation rates.  Note that this
diverges as $(v_c - v)^{-1}$ as $v \to v_c$ from below, where $v_c
\equiv 2 \sqrt{D (a_1 - b_1)}$ is the velocity at which the type-1
dominated ground state eigenvalue $\Gamma_{gs}$ reaches its
delocalization threshold $\Gamma_1^*$.  This divergence will be
difficult to observe in real biological systems because
$v_s < v_c$.

\emph{(3) The ratio of the two genotypes is proportional to
$\mu$}.  Below the switching velocity, the ratio of the number of
type $2$ individuals to type $1$ approaches $\frac{\mu_1 a_1}{a_1
- a_2}$ at long times.  Above the switching velocity, the ratio of
type $1$ to type $2$ is $\frac{\mu_2 a_2}{b_2 - b_1}$ at long
times.  This result, however, is only true within the linear
model.  If we add a saturation term to the dynamics, this term
will affect the long-time population ratios.

\emph{(4) The ratio of genotype 1 to genotype 2 is independent of
$v$ except when crossing $v_s$.} We might naively expect that as
we increase $v$, the population gets driven more towards the
desert, and thus the population shifts away from genotype $1$
towards genotype $2$.  The gauge transformation ensures that this
does not happen.  Until we cross $v_s$, the ratio of genotypes in
the ground state eigenfunction, and thus the population, remains
constant.  At $v_s$ there is a sharp transition and the ratio of
the genotypes shifts radically in favor of genotype $2$.  As we
continue to increase the velocity, the ratio again remains
constant.  This result, however, is also only true within the
linear model.  A nonlinear saturation term ``softens'' the
transition at $v_s$, as described in section \ref{nonlinear}.

\emph{(5) The ``right'' ratio of genotypes dominates exponentially.} If the
initial state of the population below the switching velocity is all type $2$,
then type $1$ will take over exponentially with a rate equal to the difference
between $\Gamma_{gs}$ and the dominant type $2$ eigenvalue.  This rate is just
$a_1 - a_2 - \mu_1 a_1 + \mu_2 a_2$, and is independent of $v$.  Similarly, if
we begin with a type-$1$ dominated population above the switching velocity, the
population will become dominated by genotype $2$ exponentially with rate $b_2 -
b_1 - \mu_2 a_2 + \mu_1 a_1$, again independent of $v$.

\subsection{Effects of a Nonlinearity} \label{nonlinear}
Thus far we have concentrated on purely linear systems, without
much discussion of nonlinear terms which cause the population to
saturate. This approximation is justified because the presence of
a nonlinear saturation term will not affect the critical
velocities. However, since all populations can be expected to
saturate at some point, it is important to consider the general
implications of such a nonlinearity.

In the discussion to this point, we have assumed that the
eigenfunction with the largest growth rate will dominate the
population, so that at long times we can neglect all but this
dominant state.  In the nonlinear case this is still roughly true.
We can account for the effects of the nonlinearity by using mode
couplings as in \cite{111}, and anticipate that the fastest
growing eigenfunction will suppress the others and dominate the
population. There is, however, one important exception to this
idea. For the problem considered here, the fastest-growing
localized eigenfunction vanishes exponentially outside of the
oasis, and hence will not suppress the growth of the
fastest-growing delocalized eigenfunction.  Thus below the
switching velocity $v_s$, the population is not in fact completely
dominated by genotype $1$. Rather, there is a genotype-$1$
dominated population inside the oasis and a genotype-$2$ dominated
delocalized population in the desert.  As we increase $v$ we
decrease the growth rate of the localized type $1$ dominated state
without changing the growth rate of the delocalized genotype $2$
dominated state. Since the overall population sizes are set by the
growth rates and the nonlinear terms, this will lead to a shift in
the overall population density from type 1 to type 2.  Thus there
will be some $v$-dependence in the genotype ratio even below
$v_s$.  Just above $v_s$, the fastest-growing delocalized
eigenfunction will not completely suppress the fastest-growing
localized eigenfunction, so similarly there will be some
$v$-dependence above $v_s$.  The transition at $v_s$ is thus
softened by the presence of the nonlinearity.  While the dominance
will still shift at this critical velocity, the transition will
have some width dependent on the saturation term.

The nonlinearity will also impose a maximum carrying capacity on
the system. As the system approaches this carrying capacity, the
growth rates will slow down.  Thus if we start with an initial
condition involving a population already near its carrying
capacity, our analysis of the rate at whch the ``right'' genotype
composition is established will be an overestimate. While the
results regarding the switching and extinction velocities and
genotype composition still hold, the dynamics of reaching the
resulting steady states will be slower. Our results for the rates
are based on the linearization of the nonlinear model around $c_1
= c_2 = 0$, and so are valid as long as the ratio of the
population to the carrying capacity is small compared to 1.

\subsection{Other Geometries} \label{geometries}
Many alternative geometries are clearly possible.  One obvious choice is a
circular oasis in an infinite desert, as considered in \cite{230}.  The
primary effect of a nontrivial two-dimensional geometry is to change the
eigenvalue spectrum.  In particular, $\Gamma_{gs}$ and, if the desert is
not infinite, $\Gamma_2^*$ will shift.  The resulting change in the
switching and extinction velocities will depend on the growth rates $a_i$
and $b_i$ and the linear size of the oasis $W$, but not on $\mu$.
However, the qualitative behavior is unaffected.  Furthermore, if $W$ is
large compared to $\sqrt{D/a}$, the typical length an individual diffuses
before giving birth (the precise requirement will vary with the specific
geometry), this shift in critical velocities will vanish and the results
will reduce to those calculated above.

\section*{Appendix A: Exact Solution for the $v=0$ System}
It is straightforward, though tedious, to find the exact solution
for the $v = 0$ eigenfunctions and eigenvalues.  We start with the
ansatz \eon \bf{\phi}^R = \left(
\begin{array}{c} A e^{-\kappa_1 x} + B e^{-\kappa_2 x} \\ C
e^{-\kappa_1 x} + E e^{-\kappa_2 x}
\end{array} \right), \eoff for $\frac{W}{2} < x$, \eon \bf{\phi}^R = \left(
\begin{array}{c} F e^{i k_1 x} + G e^{-i k_1 x} + H e^{ik_2 x} + I e^{-ik_2 x} \\
J e^{ik_1 x} + K e^{-ik_1 x} + L e^{ik_2 x} + M e^{-ik_2 x}
\end{array} \right) \eoff for $-\frac{W}{2} < x < \frac{W}{2}$, and \eon \bf{\phi}^R =
\left( \begin{array}{c} N e^{\kappa_1 x} + O e^{\kappa_2 x} \\ P
e^{\kappa_1 x} + Q e^{\kappa_2 x} \end{array} \right) \eoff for $x
< - \frac{W}{2}$, where the 16 parameters $A, B, C, E, F, G, H, I,
J, K, L, M, N, O, P$, and $Q$ are constant coefficients. Demanding
that this ansatz satisfy the eigenvalue equation yields a system
of $12$ equations relating these constant coefficients, $\kappa,
k$, and the eigenvalue $\Gamma$.  We use $8$ of these equations to
eliminate $8$ of the $16$ coefficients, and the remaining $4$ to
determine $\kappa_1, \kappa_2, k_1$, and $k_2$ in terms of the
eigenvalue $\Gamma$.  We find that for small $\mu_1$ and $\mu_2$
($\frac{\mu_1 \mu_2 b_1 b_2}{(b_1 - b_2)^2} \ll 1$ and
$\frac{\mu_1 \mu_2 a_1 a_2}{(a_1 - a_2)^2} \ll 1$),
\begin{eqnarray} D \kappa_1^2 & = & \Gamma - b_2 + \mu_2 a_2 +
\frac{\mu_1 a_1 \mu_2 a_2}{b_2 - b_1} \\ D \kappa_2^2 & = & \Gamma
- b_1 + \mu_1 a_1 - \frac{\mu_1 a_1 \mu_2 a_2}{b_2 - b_1} \\ D
k_1^2 & = & - \Gamma + a_1 - \mu_1 a_1 + \frac{\mu_1 a_1 \mu_2
a_2}{a_1 - a_2} \\ D k_2^2 & = & - \Gamma + a_2 - \mu_2 a_2 -
\frac{\mu_1 a_1 \mu_2 a_2}{a_1 - a_2}, \end{eqnarray}  which
serves to confirm our perturbation theory calculation. The
expressions when $\mu_1$ and $\mu_2$ are not small are
straightforward to calculate but unwieldy to write down.

We now demand that $\bf{\phi}$ and $\od{\bf{\phi}}{x}$ be
continuous at $x = \pm \frac{W}{2}$, yielding $8$ equations for
the remaining $9$ unknowns ($8$ constant coefficients and
$\Gamma$). These results lead to a transcendental equation for
$\Gamma$, the solutions to which are the eigenvalues of the
system. Choosing a particular eigenvalue from among these possible
solutions, we can easily determine the remaining unknowns (up to
an overall normalization) from the rest of the equations.  By
requiring \eon \int \bf{\phi}^L(x) \cdot \bf{\phi}^R(x) dx = 1,
\eoff we determine the normalization, and thus the exact solution.

In practice, the transcendental equation for $\Gamma$ is quite
complicated and we can only solve it numerically.  However, this
exact approach does provide a useful check to the discretized
numerical solution described in section \ref{simulation}.  The
most important result is the $\mu = 0$ ground state eigenvalue
$\Gamma^{\mu=v=0}_{gs}$.  Provided that the oasis is much wider
than the distance a typical individual diffuses before giving
giving birth, we have $\Gamma_{gs}^{\mu=v=0} \approx a_1$
\cite{230}.

\section*{Appendix B: Perturbation Theory in $\mu_1$ and $\mu_2$}
We can use standard perturbation theory from quantum mechanics
\cite{qmbook} to determine the results for $\mu_1, \mu_2 > 0$ from
the $\mu_1 = \mu_2 = 0$ solution.  We first rewrite the Liouville
operator as $\mat{L}_{v=0} = \mat{L}_0 + \mat{L}_1$, where
$\mat{L}_1$ is proportional to the (small) mutation rates $\mu_1$
and $\mu_2$, and \begin{eqnarray} \mat{L}_0 & = & \left(
\begin{array}{cc} D \partial_x^2 + r_1(x) & 0 \\ 0 & D
\partial_x^2 + r_2(x)
\end{array} \right), \\ \mat{L}_1 & = & \left( \begin{array}{cc}
- \mu_1 a_1 & \mu_2 a_2 \\ \mu_1 a_1 & - \mu_2 a_2 \end{array}
\right) , \end{eqnarray} where $r_i(x) = (a_i + \mu_i a_i) \theta
(W - |x|) + (b_i + \mu_i a_i ) \theta (|x| - W)$.  We know the
eigenvalues and eigenfunctions of $\mat{L}_0$ from section
\ref{analysis} and from \cite{230}.  For our purposes, all that is
necessary is that the eigenfunctions are an orthonormal set of
left and right functions $\bf{\phi}_n^L$ and $\bf{\phi}_n^R$.

From non-degenerate time-independent perturbation theory we find
that the $\mu > 0$ eigenvalues are related to the $\mu = 0$
eigenvalues by the formula {\small \begin{eqnarray} \Gamma_n & = &
\Gamma_n^{\mu = 0} +
\int \bf{\phi}_n^{L}(x) \mat{L}_1 \bf{\phi}_n^R(x) dx + \\
& & + \sum_{m \neq n} \frac{ \left( \int \bf{\phi}_m^{L}(x)
\mat{L}_1 \bf{\phi}_n^R(x) dx \right)  \left( \int
\bf{\phi}_n^{L}(x) \mat{L}_1 \bf{\phi}_m^R(x) dx
\right)}{\Gamma_n^{\mu = 0} - \Gamma_m^{\mu=0}}, \nonumber
\end{eqnarray} } plus terms of third and higher order in $\mat{L}_1$
\cite{qmbook}. Note that this result differs slightly from
standard theory because $\mat{L}_1$ is non-Hermitian.  The
first-order term in $\mat{L}$ is straightforward to calculate.  It
is simply $- \mu_i a_i$ for states dominated by genotype $i$. The
second-order term is more complicated, because the $\mu = 0$
eigenfunctions are of the form \eon \bf{\phi}_n^R = \left(
\begin{array}{c} \psi_n^1 \\ 0 \end{array} \right) \quad
\mathrm{or} \quad \bf{\phi}_n^R = \twovect{0}{\psi_n^2}, \eoff
where the $\{ \psi_n^1 \}$ and $\{ \psi_m^2 \}$ each form an
orthonormal set of eigenfunctions of the one-genotype problem. The
$\psi_n^1$ and $\psi_m^2$ are almost, but not quite, orthonormal
to each other.

In the approximation that these two sets of eigenfunctions are
indeed orthonormal, the second order term for $\Gamma_n$ is easy
to calculate.  For genotype-1 dominated localized states, it is
$\frac{\mu_1 a_1 \mu_2 a_2}{a_1 - a_2}$.  For genotype-2 dominated
delocalized states, it is $\frac{\mu_1 a_1 \mu_2 a_2}{b_2 - b_1}$.
In the same approximation, the corrections to the eigenfunctions
are given by \eon \bf{\phi}_n^{1, v=0, R} = \left(
\begin{array}{c} \psi_n^1 \\ \frac{\mu_1 a_1}{a_1 - a_2} \psi_n^2
\end{array} \right), \eoff for a genotype-1 dominated localized
state, and \eon \bf{\phi}_n^{2, v=0, R} = \left(
\begin{array}{c} \frac{\mu_2 a_2}{b_2 - b_1} \psi_n^1 \\ \psi_n^2 \end{array}
\right) \eoff for a genotype-2 dominated delocalized state.

These results, and the observation that $\Gamma_{gs}^{\mu = v = 0}
= a_1$ and $\Gamma_2^{*, \mu = v = 0} = b_2$, allow us to
calculate the critical velocities.  To second order in $\mu_1$ and
$\mu_2$, the extinction velocity is given by \eon v_e = 2 \sqrt{(D
a_1)} \left[ 1 - \frac{\mu_1}{2} - \frac{\mu_1^2}{8} + \frac{\mu_1
\mu_2 a_2}{2(a_1 - a_2)} \right], \eoff and the switching velocity
is \begin{eqnarray} v_s & = & 2 \sqrt{D (a_1 - b_2)} \left[ 1 -
\frac{\mu_1 a_1 - \mu_2 a_2}{2(a_1- b_2)} - \frac{(\mu_1 a_1 -
\mu_2 a_2)^2}{8 (a_1 - b_2)^2} \right. \nonumber \\ & & \left. +
\frac{\mu_1 a_1 \mu_2 a_2}{2(a_1 - a_2) (a_1 - b_2)} - \frac{\mu_1
a_1 \mu_2 a_2}{2(b_2 - b_1)(a_1-b_2)} \right] .
\end{eqnarray}  The first order part of this result is quoted in
section \ref{analysis} above.

This perturbation expansion relies on the assumption that $\mu_1$ and
$\mu_2$ are small.  Specifically, we require \eon \frac{\mu_i a_i}{a_2 -
a_1} \ll 1 \qquad \frac{\mu_i a_i}{b_2 - b_1} \ll 1, \eoff for $i = 1, 2$
for all of these results to be valid.

\section*{Appendix C: Lattice Approximation}
Here we describe the details of the lattice approximation to the
Liouville operator.  We begin by discretizing space, replacing it
by a regular lattice with lattice constant $\ell_0$.  We define
$c_x^\alpha$ to be the population of genotype $\alpha$ at the
lattice point $x$.  We then have to solve the eigenvalue equation
\eon \od{c_x^\alpha (t)}{t} = \sum_{x', \alpha'} \mat{L} (x,
\alpha, x', \alpha') c_{x'}^{\alpha'} (t) = \Gamma c_x^\alpha (t).
\eoff The discretized version of the Liouville operator is
\begin{eqnarray} \mat{L} & = & \frac{D}{\ell_0^2} \sum_x
\sum_{\alpha=1}^2 \left[ e^{-\frac{v \ell_0}{2D}} | x
\rangle^\alpha \chem{\langle}{\alpha} x + \ell_0| \right.
\nonumber \\ & & \left. + e^{\frac{v \ell_0}{2D}} | x + \ell_0
\rangle^\alpha \chem{\langle}{\alpha} x| - 2 \cosh \left( \frac{v
\ell_0}{2D} \right) |
x \rangle^\alpha \chem{\langle}{\alpha} x | \right] \nonumber \\
& & +\sum_x \left[ U_1(x) |x \rangle^1 \chem{\langle}{1} x | +
U_2(x) |x \rangle^2 \chem{\langle}{2} x | \nonumber \right. \\
& & +\left. \mu_1 a_1 |x \rangle^1 \chem{\langle}{2} x | + \mu_2
a_2 |x \rangle^2 \chem{\langle}{1} x | \right], \end{eqnarray}
where we have used the notation $|x \rangle^{\alpha}
\chem{\langle}{\beta} y |$ to mean the tensor product of localized
states corresponding to $c_x^\alpha$ and $c_y^\beta$. We define
$U_1(x) = (a_1 - \mu_1 a_1) \theta (W - |x|) + (b_1 - \mu_1 a_1)
\theta (|x| - W),$ and $U_2(x) = (a_2 - \mu_2 a_2) \theta (W -
|x|) + (b_2 - \mu_2 a_2) \theta (|x| - W)$. We impose a finite
size on the system $L$ (with periodic boundary conditions) and a
finite width of the oasis $W$, with $W \ll L$.

In order for the lattice approximation to be valid, the lattice
must be fine enough that variations in the eigenfunction $\phi$
between lattice points is small.  This means we must require
$\frac{\ell_0 | \partial_x \phi_n^{R, \alpha}(x)| }{\phi_n^{R,
\alpha}(x)} \ll 1$.  For small $v$ this reduces to
$\kappa_{\alpha}^n \ell_0 \ll 1$, $k_{\alpha}^n \ell_0 \ll1$, and
for large $v$ we need $\frac{v \ell_0}{2D} \ll 1$ \cite{230}.
These conditions are satisfied for all of the calculations
discussed here.

It is now straightforward to numerically solve for the eigenvalues
and eigenstates of the lattice version of the Liouville operator
$\mat{L}$. The results quoted here all use a system size $L = 512
\ell_0$, with an oasis width $W = 10 \ell_0$ and periodic boundary
conditions.

In comparing results, it is useful to shift to dimensionless
units.  We define a dimensionless velocity \eon \bar v =
\frac{v}{2 \sqrt{a_1 D}}. \eoff  Note that for $\mu_1 = \mu_2 =
0$, the extinction velocity $\bar v_e = 1$.  We then scale all the
growth rates to $a_1$ by defining \mon \bar a_1 = \frac{a_1}{a_1}
= 1, \quad \bar b_1 = \frac{b_1}{a_1}, \quad \bar a_2 =
\frac{a_2}{a_1}, \quad \bar b_2 = \frac{b_2}{a_1}, \quad \bar
\Gamma = \frac{\Gamma}{a_1}. \moff  This implies that the
dimensionless diffusion coefficient is \eon \bar D = \frac{1}{4}.
\eoff The mutation rates are already dimensionless.  We use these
redefined units in \fig{fig2}, \fig{fig22}, and \fig{fig3}.

\section*{Appendix D: Discrete Simulation}
Here we describe the details of the discrete, individual-based
simulations.  We begin with a discretized spatial lattice
containing a uniform distribution of individuals of both types. We
also discretize time, dividing it into small intervals of size
$\Delta t$.  At each time, we select a lattice point at random.
The individuals at this point can give birth, move due to
diffusion or drift, or mutate with appropriate probabilities.  The
probabilities used are simply the coefficients of the off-diagonal
elements of the discretized Liouville operator described in
Appendix C, times $\Delta t$. We set $\Delta t$ to be sufficiently
small that the probability of two or more events per step is
negligible.  We impose a saturation effect (analagous to a
nonlinear term in \eq{eq4}) by setting a maximum number of
individuals per spatial point.

For the simulations described in \fig{fig3}, we use the parameters
$\bar a_1 = 1, \bar a_2 = 0.6, \bar b_1 = -1, \bar b_2 = -0.6$,
and $\bar D = 0.25$, where the overbars denote the dimensionless
parameters defined in Appendix C. We use a lattice with $512$
points, with an oasis of width $10$ points and periodic boundary
conditions, and a maximum of $200$ individuals per point.

For these parameter values, $v_s$ is impossible to determine
because the populations are extinct at such high velocities.
However, we can determine $v_e$ as a function of $\mu_1$ and
$\mu_2$.  For each value of $\mu_1$ and $\mu_2$ that we test,
we plot the total number of individuals in the steady state
distribution as a function of $v$.  This exhibits a transition
from some maximum number of individuals for small $v$ to
approximately $0$ individuals for large $v$. Because of the
imposed saturation effect, the transition is not perfectly sharp
but rather has some small width. We define the transition to occur
at the point at which the number of individuals has dropped to
$\frac{1}{10}$ the number for $v=0$. Making a different definition
would shift the extinction velocities slightly.

We have also run simulations for parameter values where $v_s$ is biologically
relevant.  From the $v$-dependence of the ratio of the number of individuals
of type $1$ to type $2$, we can determine the value of $v_s$ in these cases.
Although the presence of saturation broadens the transition as in the case of
$v_e$, the results match our analytical predictions.


\begin{thebibliography}{99}
\bibitem{252}
Hatano, N and D. R. Nelson (1998).  ``Non-Hermitian Delocalization and
Eigenfunctions.''  \emph{Physical Review B} \textbf{58}: 8384.
\bibitem{255}
Hatano, N and D. R. Nelson (1997).  ``Vortex Pinning and Non-Hermitian
Quantum Mechanics.''  \emph{Physical Review B} \textbf{56}: 8651.
\bibitem{111}
Nelson, D. R. and N. Shnerb (1998).  ``Non-Hermitian Localization and
Population Biology.''  \emph{Physical Review E} \textbf{58}: 1383.
\bibitem{230}
Dahmen, K. A., D. R. Nelson, and N. Shnerb (2002).  ``Life and Death Near
a Windy Oasis.''  \emph{Journal of Mathematical Biology} \textbf{41}: 1.
\bibitem{231}
Neicu, T., A. Pradhan, D. A. Larochelle, and A. Kudrolli (2000).
``Extinction Transition in Bacterial
Colonies Under Forced Convection.''  \emph{Physical Review E} \textbf{62}:
1059.
\bibitem{245}
Altmeyer, S. and J. S. McCaskill (2001).  ``Error Threshold for Spatially
Resolved Evolution in the Quasispecies Model.''  \emph{Physical Review
Letters} \textbf{86}: 5819.
\bibitem{227}
Gerland, U. and T. Hwa (2002).  ``On the Selection and Evolution of
Regulatory DNA Motifs.''  \emph{Journal of Molecular Evolution}
\textbf{55}: 386.
\bibitem{altmeyerref1}
Eigen, M (1971).  \emph{Naturwissenschaften} \textbf{58}: 465.
\bibitem{altmeyerref2}
Eigen, M, J. S. McCaskill, and P. Schuster (1989).  \emph{Advances in
Chemical Physics} \textbf{75}: 149.
\bibitem{altmeyerref5}
McCaskill, J. S. (1984).  \emph{Journal of Chemical Physics} \textbf{80}:
5194.
\bibitem{46}
Cardy, J. (2001). ``Renormalisation Group Approach to
Reaction-Diffusion Problems.''  http://xxx.lanl.gov/list/cond-mat,
paper number 9607163v2.
\bibitem{54}
Shnerb \emph{et. al.} (2000).  ``Adaptation of Autocatalytic Fluctuations
to Diffusive Noise.''  http://xxx.lanl.gov/list/cond-mat, paper number
0007097.
\bibitem{56}
Bettelheim, E., O. Agam, and N. Shnerb (2000).  ``Quantum Phase
Transitions in Classical Nonequilibrium Processes.''
http://xxx.lanl.gov/list/cond-mat, paper number 9908450v4.
\bibitem{58}
Cardy, J. and U. Tauber (1996).  ``Theory of Branching and Annihilating
Random Walks.'' \emph{Physical Review Letters} \textbf{77}:4780.
\bibitem{59}
Doi, M. (1976).  ``Second Quantization Representation for Classical
Many-Partical System.''  \emph{J. Phys. A} \textbf{9}:1465.
\bibitem{60}
Doi, M. (1976).  ``Stochastic Theory of Diffusion-Controlled Reaction.''
\emph{J. Phys. A.} \textbf{9}:1479.
\bibitem{61}
Peliti, L. (1985).  ``Path Integral Approach to Birth-Death Processes on a
Lattice.''  \emph{J. Physique} \textbf{56}:1469.
\bibitem{endnote2}
All of the eigenfunctions other than the ground state involve
negative (and sometimes complex) values of the population density
at certain points.  Only the ground state is real and positive
everywhere.  Nevertheless, these eigenfunctions can combine to
produce real nonnegative distributions, and certain relations
among them ensure that if the population is initially everywhere
real and nonnegative it will remain so for all time.  However, all
such combinations must include the ground state.
\bibitem{endnote4}
Note that what we refer to as the ``ground state'' is actually the
ground state (whose eigenvalue has the smallest real part) of the
operator $- \mat{L}$.
\bibitem{qmbook}
Landau, L. D. and E. M. Lifshitz (1991).  \emph{Quantum Mechanics
(Non-Relativistic Theory)}.  Pergamon Press.
\bibitem{endnote3}
This approximation can be relaxed following \cite{qmbook}, giving a
$\Gamma_{gs}$ that depends on the growth rates, $D$, and the width of the oasis.  This result is
highly dependent on the particular geometry of the oasis, however, and
always reduces to our approximation provided the oasis is sufficiently large.
\bibitem{endnote1}
This is easiest to see in the case $\mu_2 = 0$.  In
this case, the critical velocity occurs when $\Gamma_{gs}$ falls below the
highest growth rate state of the genotype-2 dominated states.  This occurs
at $\mu_1 = \frac{a_1 - a_2}{a_1}$, which is identical to the non-spatial
calculation of the error threshold.  However, this value for $\mu$
explicitly violates our perturbation theory condition that $\frac{\mu_1
a_1}{a_1 - a_2} \ll 1$.  Thus the requirement that our perturbation theory
analysis be valid is equivalent to saying that we must be well below the
mutational error threshold.  It would in principle be possible to use the
exact solution discussed in Appendix A to explore the velocity-driven
transition near the mutational threshold.
\end{thebibliography}
\end{document}